\documentclass[DIV=calc,paper=a4,fontsize=11pt,twocolumn]{scrartcl} 

\pdfoutput=1
\pdfmapfile{+txfonts.map}

\usepackage[english]{babel}
\usepackage[protrusion=true,expansion=true]{microtype}
\usepackage{amsmath,amsfonts,amsthm}
\usepackage[final]{graphicx}
\usepackage{xcolor}
\usepackage[normal,small,hypcap,up,labelfont=bf,textfont=it]{caption}
\usepackage{subfig}
\usepackage{booktabs}
\usepackage{fix-cm}
\usepackage{amssymb,amsfonts}
\usepackage{dsfont}
\usepackage{bbm}
\usepackage{cite}
\usepackage[utf8]{inputenc}
\usepackage[perpage,symbol*]{footmisc}
\usepackage[varg]{txfonts}
\usepackage{balance}
\usepackage{fancyhdr}
\PassOptionsToPackage{hyphens}{url}\usepackage[pdfencoding=auto,psdextra]{hyperref}
\usepackage{bookmark}
\usepackage{verbatim}
\usepackage{fontenc}
\usepackage{cuted}

\DeclareCaptionFont{mycolor}{\color[HTML]{000000}}
\usepackage{sectsty}													
\allsectionsfont{
\color[HTML]{31ADF3}\usefont{OT1}{phv}{b}{n}
}

\sectionfont{
\color[HTML]{31ADF3}\usefont{OT1}{phv}{b}{n}
}

\usepackage{fancyhdr}												
\usepackage{lettrine}
\newcommand{\initial}[1]{%
\lettrine[lines=3,lhang=0.3,nindent=0em]{
\color[HTML]{31ADF3}
{\textsf{#1}}}{}}

\usepackage{titling}															

\newcommand{\HorRule}{\color[HTML]{31ADF3}
\rule{\linewidth}{1pt}%
}

\pretitle{\vspace{-30pt} \begin{flushleft} \HorRule
\fontsize{34}{34} \usefont{OT1}{phv}{b}{n} \color[HTML]{31ADF3} \selectfont
}
\title{Is the Quantum State Real in the Hilbert Space Formulation?}					
\posttitle{\par\end{flushleft}\vskip 0.5em}

\preauthor{\begin{flushleft}\large \lineskip 0.5em \usefont{OT1}{phv}{b}{sl} \color[HTML]{31ADF3}}
\author{Mani L. Bhaumik\\[8pt]}											
\postauthor{\footnotesize \usefont{OT1}{phv}{m}{sl} \color[HTML]{000000}
Department of Physics and Astronomy, University of California, Los Angeles, USA. E-mail: \href{mailto:bhaumik@physics.ucla.edu}{bhaumik@physics.ucla.edu}\\[10pt]		
\scriptsize\usefont{OT1}{phv}{m}{n} \color[HTML]{31ADF3}{\textbf{Editors: \emph{Zvi Bern} \& \emph{Danko Georgiev}} }\\[5pt]
\color[HTML]{000000}{Article history: Submitted on November 5, 2020;  Accepted on December 18, 2020; Published on December 19, 2020.}
\par\end{flushleft}\HorRule}

\date{}																				

\begin{document}
\maketitle
\thispagestyle{fancy} 			
\initial{T}\textbf{he persistent debate about the reality of a quantum state has recently come under limelight because of its importance to quantum information and the quantum computing community. Almost all of the deliberations are taking place using the elegant and powerful but abstract Hilbert space formalism of quantum mechanics developed with seminal contributions from John von Neumann. Since it is rather difficult to get a direct perception of the events in an abstract vector space, it is hard to trace the progress of a phenomenon. Among the multitude of recent attempts to show the reality of the quantum state in Hilbert space, the Pusey--Barrett--Rudolph theory gets most recognition for their proof. But some of its assumptions have been criticized, which are still not considered to be entirely loophole free. A straightforward proof of the reality of the wave packet function of a single particle has been presented earlier based on the currently recognized fundamental reality of the universal quantum fields. Quantum states like the atomic energy levels comprising the wave packets have been shown to be just as real. Here we show that an unambiguous proof of reality of the quantum states gleaned from the reality of quantum fields can also provide an explicit substantiation of the reality of quantum states in Hilbert space.\\ Quanta 2020; 9: 37--46.}

\begin{figure}[b!]
\rule{245 pt}{0.5 pt}\\[3pt]
\raisebox{-0.2\height}{\includegraphics[width=5mm]{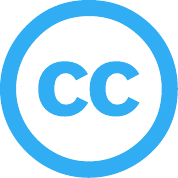}}\raisebox{-0.2\height}{\includegraphics[width=5mm]{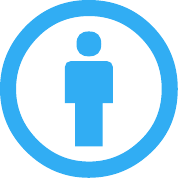}}
\footnotesize{This is an open access article distributed under the terms of the Creative Commons Attribution License \href{http://creativecommons.org/licenses/by/3.0/}{CC-BY-3.0}, which permits unrestricted use, distribution, and reproduction in any medium, provided the original author and source are credited.}
\end{figure}

\section{Introduction}

The debate about the reality of quantum states is as old as quantum
physics itself. The objective reality underlying the manifestly bizarre
behavior of quantum objects is conspicuously at odds with our daily
classical physical reality. The scientific outlook of objective reality
commenced with the precepts of classical physics that cemented our
notion of reality for centuries. Discoveries starting in the last
decade of the nineteenth century revealing the uncanny quantum world
in the microscopic domain shook that perception.

With the particular exception of Einstein, who was the lone supporter
of his postulated wave-particle duality for photons for about two
decades, physicists continued to think of the microscopic quantum
world within the confines of the ingrained classical physics. Finally,
with de Broglie's proposed extension of the wave--particle duality
to matter particles like electrons and its experimental verification,
irrevocably opened the door ushering in the bizarre new world of quantum
physics.

With some talented younger physicists like Schr\"{o}dinger, Heisenberg,
Born, Dirac and others, the development of the quantum physics proceeded
in a break neck speed starting in early 1926. Although these efforts
led to the most successful description of events in the atomic domain,
the revelations of quantum physics were so weird that immediately
a debate started about the significance of it all.

Soon the dispute came to a climax at the 1927 Solvay Conference in
Belgium with the famed Bohr--Einstein debate. While Bohr insisting
that there was no reality before a quantum state is measured, Einstein
maintained there must be a reality even before a quantum state is
observed. Almost a century later the debate is surprisingly still
thriving. A conspicuous example is the substantial account presented
in the recent review article by Matthew Leifer \cite{key-1}. All
of these contemporary considerations are conducted using the abstract
Hilbert space formulation of quantum mechanics initiated by John von
Neumann.

After John Bell's epochal paper \cite{key-2} presented Bell's inequality
and its numerous experimental substantiations, the reality of the
quantum state is now more acceptable in contrast to the conclusion
of earlier Bohr--Einstein debate. The most prominent recent theory
of reality is presented by Matthew Pusey, Jonathan Barrett, and Terry
Rudolph \cite{key-3}. Other theories are also advanced by Lucien
Hardy \cite{key-4} as well as Roger Colbeck and Renato Renner \cite{key-5}.
However, none of these latest advances is considered to be entirely
loophole free. Leifer's considerably extensive review \cite{key-1}
provides a distinct example of the difficulties of reaching a definitive
conclusion using the circuitous way of deliberations in the abstract
Hilbert space. Here we present a rather straightforward way to prove
the reality of the quantum state.

In order to avoid any possible confusion, it would be prudent to agree
upon the definition of reality. In this regard, we rely upon the generally
acknowledged connotation of reality. We consider something to be physically
real if it is independently observed by several people and they agree
with each other that the result of their observations is the same.
Accordingly, one could rely on the following notions of the distinguished
contemporary physicists for our understanding of reality.

Referring to the outstanding developments in the cutting-edge quantum
field theory or QFT in short, the distinguished Physics Nobel Laureate
Frank Wilczek asserts
\begin{quote}
the standard model is very successful in describing reality---the
reality we find ourselves inhabiting. \cite[p.~96]{key-6}
\end{quote}
Wilczek additionally enumerates
\begin{quote}
The primary ingredient of physical reality, from which all else is
formed, fills all space and time. Every fragment, each space-time
element, has the same basic properties as every other fragment. The
primary ingredient of reality is alive with quantum activity. Quantum
activity has special characteristics. It is spontaneous and unpredictable.
\cite[p.~74]{key-6}
\end{quote}
Another esteemed Physics Nobel Laureate Steven Weinberg confirms
\begin{quote}
the Standard Model provides a remarkably unified view of all types
of matter and forces (except for gravitation) that we encounter in
our laboratories, in a set of equations that can fit on a single sheet
of paper. We can be certain that the Standard Model will appear as
at least an approximate feature of any better future theory. \cite{key-7}
\end{quote}
Thus, it would be cogent to consider the space filling universal Quantum
Fields as the primary ingredients of physical reality uncovered by
us so far. An abundant proof of this can be encountered all around
us in several different ways. The most direct convincing evidence
comes from the fact that elementary particles like an electron has
exactly the same properties, such as mass-energy, charge, spin etc.,
irrespective of when or where in the universe it comes into existence---in
the big bang, in astrophysical processes throughout the eons or anywhere
in a lab in the world.

A manifestation of the fluctuations of the quantum fields in a phenomenon
like the electron anomalous $g$-factor agrees up to an unprecedented
twelve decimal places when the experimental results are compared to
the theoretical computation. Observed phenomena like Lamb shift, Casimir
effect further assert the existence of the fluctuations of the quantum
fields. A very dramatic confirmation of the indispensable effects
of the quantum field fluctuations comes from the mass of the composite
particles like protons and neutrons. The mass of the three valence
quarks in a proton provided by the Higgs boson is only about $9$
Mev while the total proton mass is a whopping $938$ Mev. This magical
``mass without mass'' ascends from the endowment of quantum fluctuations.

Perhaps the most spectacular graphic evidence is provided by the observed
anisotropy in the cosmic microwave background radiation with their
presumed origin in the cosmic inflation in the early universe when
the quantum fluctuations of the reputed inflaton field enormously
expanded from the very microscopic to macroscopic dimensions providing
seeds for galaxy formation afterwards. Any reasonable concept of physical
reality should then owe its eventual origin to the fundamental reality
of quantum fields and their characteristic attributes.

The elementary particles like electrons, one of the members of the
initial act of material formation from the abstract but physical quantum
fields, are quanta of the fields. Each of them can be rendered as
a wave packet consisting of an admixture of the various fields. Accordingly,
the wave packet function of the elementary particle ought to be considered
as real as the primary quantum fields. More generally the fields whose
quantization produces the 24 other observed elementary particles in
Nature, such as the photons, $W^{+}$, $W^{-}$, $Z^{0}$, quarks
and gluons, embodied in the Standard Model of particle physics should
be just as real. Consequently so would be the composite particles
like protons and neutrons. We can then consider the quantum states
of atoms and molecules comprising the elementary and composite particles
real as well.

This would empower us with confidence to extend the reality of the
quantum states to those in Hilbert space formalism including all its
operations to be as real as the wave packet of a single quantum particle
and the wave functions of all quantum states of the Schr\"{o}dinger and
Heisenberg formulation.

\section{The wave function of an electron}

A particle like an electron arises as a quantized excitation of the
underlying electron quantum field. Such an energy-momentum eigenstate
of the field can be expressed as a specific Lorentz covariant superposition
of field shapes of the electron field along with all the other quantum
fields of the Standard Model of particle physics.

Superposition of field shapes in a one-particle state evolves in a
simple wavelike manner with time dependence $e^{-\imath\omega t}$.
The individual field shapes, each with their own computable dynamic
time evolution, are actually the vacuum fluctuations comprising the
very structure of the energy-momentum eigenstate. The quantum fluctuations
are evanescent in the sense that they pass away soon after coming
into being. But new ones are constantly boiling up to establish an
equilibrium distribution so stable that their contribution to the
electron $g$-factor, as mentioned earlier, results in a measurement
accuracy of one part in a trillion \cite{key-8}.

The Lorentz covariant superposition of fluctuations of all the quantum
fields in the one-particle quantum state can be conveniently depicted
leading to a well behaved smooth wave packet that is everywhere continuous
and continuously differentiable.

However, the detailed quantitative calculations are quite involved.
For simplicity, we can get the result using a heuristic perspective.
For this purpose, let us consider an isolated single quantum of a
non-interacting electron quantum field. Since no force is acting on
such an electron, its momentum would be constant and therefore its
position would be indefinite since a regular ripple from a free electron
field with a very well-defined energy and momentum is represented
by a delocalized periodic function.

Recalling that the electron in reality is an admixture of all the
quantum fields of the standard model, it should be noted that in the
non-relativistic regime, there would not be enough energy to create
any new particle. Consequently, the contribution of the different
quantum fields to the single particle would comprise of irregular
disturbances of the fields with energy off the mass shell, resulting
in the electron ripple to be very highly distorted.

It is well known that such a pulse, no matter how deformed, can be
expressed by a Fourier integral with weighted linear combinations
of simple periodic wave forms like trigonometric functions \cite{key-9}.
The result would be a wave packet function to represent an electron
that would embody a fundamental reality of the universe since all
the amplitudes of the wave packet would consist of contributions of
irregular disturbances of the various real primary quantum fields.

The wave function $\psi(x)$, for simplicity in one dimension, will
be given by the inverse Fourier transform
\begin{equation}
\psi(x)=\frac{1}{\sqrt{2\pi}}\int_{-\infty}^{+\infty}\tilde{\psi}\left(k\right)e^{\imath kx}dk\label{eq:1}
\end{equation}
where $\tilde{\psi}(k)$ is a function that quantifies the amount
of each wave number component $k=\frac{2\pi}{\lambda}$ that gets
added to the combination.

From Fourier analysis, we also know that the spatial wave function
$\psi(x)$ and the wave number function $\tilde{\psi}(k)$ constitute
a Fourier transform pair. Therefore, we can find the wave number function
through the forward Fourier transform as
\begin{equation}
\tilde{\psi}\left(k\right)=\frac{1}{\sqrt{2\pi}}\int_{-\infty}^{+\infty}\psi\left(x\right)e^{-\imath kx}dx\label{eq:2}
\end{equation}
Thus, the Fourier transform relationship between $\psi(x)$ and $\tilde{\psi}(k)$,
where $x$ and $k$ are known as conjugate variables, can help us
determine the frequency or the wave number content of any spatial
wave function.

So far, we have considered only a fixed momentum. Obviously in any
dynamical system, momentum would be expected to change all the time.
As the momentum increases, the shape of the curve representing equation
(\ref{eq:2}) in the momentum space would be taller and thinner while,
being a conjugate variable, the curve in the position space would
be shorter and wider. This naturally leads us to the postulate of
famed uncertainty principle, which is evidently a natural consequence
of the position and momentum being conjugate variables in wave packets.

Fortunately, a fairly rigorous underpinning of the wave packet for
a single particle QFT state in position space for a scalar quantum
field has been provided by Robert Klauber \cite[p.~275]{key-10}.
Since particles of all quantum fields are invariably an admixture
of contributions from essentially all the fields of the Standard Model,
the wave packet function of a single particle of a scalar quantum
field can be considered to be qualitatively representative of those
of the vector fields like the electron quantum field.

The mathematical form in position space of a wave packet function
of a particle for a scalar field is \cite{key-10}
\begin{equation}
|\phi\rangle=\left|\int\left(2\pi\right)^{-\frac{3}{2}}A\left(k'\right)e^{-\imath k'x}\,d^{3}k'\right\rangle \label{eq:3}
\end{equation}
Using the more common convention of $e^{\imath k'x}$ in position
space for an inverse Fourier transform, $k$ instead of $k'$ for
momentum, $\psi(x)$ for $|\phi\rangle$ and $\tilde{\psi}(k)$ for
$A(k')$, equation (\ref{eq:3}) in one space dimension becomes exactly
equation (\ref{eq:1}), thus confirming that the single particle wave
packet function given by (\ref{eq:1}) can be rigorously derived from
QFT.

In order to determine the time evolution of the wave packet function,
we need to incorporate the time term to the spatial function. Accordingly,
\begin{equation}
\psi(x,t)=\frac{1}{\sqrt{2\pi}}\int_{-\infty}^{+\infty}\tilde{\psi}\left(k\right)e^{i\left(kx-\omega t\right)}dk\label{eq:4}
\end{equation}
where $\omega=2\pi\nu$ is the angular frequency.

On a cursory glance, the wave packet in equation (\ref{eq:4}) looks
very similar to the one that has always been used in quantum mechanics
so far, for example, to solve Schr\"{o}dinger equation for the hydrogen
atom. But in reality the wave function given by (\ref{eq:4}) is substantially
different in character since the amplitudes are fundamentally real
physical entities owing their origin to the primary reality of different
quantum fields.

When quantum mechanics, with its essential requirement of a wave packet
function to represent a localized particle, emerged in the atomic
domain of physics and commenced explaining its mysterious accomplishments
with uncanny consistency, it appeared totally contrary to our intuition
developed from classical physics. Naturally, the wave packet of a
particle was considered to be just a fictitious mathematical construct
merely necessary for carrying out calculations.

At best, the amplitudes of a wave packet are considered to be real
probability amplitudes to find the particle during a measurement.
This is consistent with the fact that the amplitudes of the wave packet
represent distinctly different properties of various quantum fields
except energy. The energy represented by any amplitude of the wave
packet has the same characteristics even though the other properties
of the amplitude may have different attributes corresponding to their
origin in the various fields.

Although the amplitude of the wave function is complex-valued, its squared modulus is real-valued.
Following Einstein, Born \cite{key-11}
had interpreted the square of the modulus of the wave amplitude as
probability for the occurrence of the particle. Thus probability $p(x)$
of finding the particle at the position $x$ in the interval $x$
and $x+dx$ is
\begin{equation}
p(x)=\psi^{*}\left(x,t\right)\psi\left(x,t\right)dx
\end{equation}

Even a century later, quantum mechanics still perplexes most people
including many scientists. This is consistent with the fact that on
an average at least one popular book on the riddles of quantum mechanics
is still being published every year by notable authors. However, as
has been explained in detail \cite{key-12}, there appear to be plausible
answers to the enigmas of quantum mechanics derived from the discoveries
of the quantum field theory of the standard model.

For example, a wave packet function is localized and therefore can
represent a quantum particle, but just holistically, since only the
totality of the wave packet represents all the conserved quantities
of the energy-momentum eigenstate of a particle such as mass, charge,
and spin. This particularly important fact requires the total wave
function to collapse during measurement, which has been one of the
most puzzling aspects of quantum mechanics. It can now be evident
in fact as a requirement because of the particular factual nature
of the wave packet.

Most of the topics in quantum mechanics, extensively used in many
diverse fields like chemistry, biology, material science, quantum
information science, involve non-relativistic quantum mechanics since
the particle speeds are lower than the speed of light. The very extensively
studied quantum objects are the hydrogen atom and the quantum harmonic
oscillator.

In the hydrogen atom, the electron revolves in the central field of
the proton producing discrete energy levels that correspond to characteristics
standing wave patterns called orbitals. These are calculated by using
the time-independent Schr\"{o}dinger equation using the real electron
wave function in equation (\ref{eq:1}). The normalized hydrogen wave
functions, using polar coordinates for convenience, can be found in
textbooks on quantum mechanics for example in \cite[\S~4.2]{key-13}.

The foremost aspect to underscore here is that the standing wave patterns
representing the quantum states of the hydrogen atom are embodiments
of the wave packet function in equation (\ref{eq:1}) that is based
on the primary reality of the quantum fields. These standing wave
patterns of the orbitals representing the quantum states are thus
objectively as real as the primary quantum fields.

In fact, reality of some of these orbitals has been established by
experimental measurements \cite{key-14}. Consequently it would be
compelling to infer that all quantum states of quantum mechanics are
based on reality, which can be extended to the quantum states described
in the abstract Hilbert space formalism, the brain child of David
Hilbert and John von Neumann.

\section{Emergence of Hilbert space}

Although von Neumann was an exceptionally talented polymath, his contribution
to the development of the Hilbert space formalism is one of his outstanding
legacies to science. Following Hilbert's preliminary thoughts, he
not only initiated the formulation but also coined the name Hilbert
space providing an advanced comprehensive mathematical foundation
of quantum theory.

\subsection{Von Neumann's contribution}

Shortly after receiving his PhD in Physics from the University of
Budapest in 1926, von Neumann decided to work under David Hilbert,
likely the foremost mathematician of the time, for completing his
habilitation program. At the University of G\"{o}ttingen, considered to
be the premiere center for mathematics in the world for quite a while,
Hilbert became interested in the indispensable application of intricate
mathematics in the emerging topics like relativity and quantum physics.
Perhaps with that perception, he was delighted to attract this mathematical
genius to his group and secured for him a six-month fellowship from
the Rockefeller financed International Education Board with additional
letters of recommendation from two of his former students, Richard
Courant and Hermann Weyl.

With all the exhilaration of the emergence of quantum physics in 1926,
Hilbert opted to devote his long-standing winter lectures on mathematics
that year to the novel topic and convinced von Neumann to join the
effort under his guidance. These impressive lectures covered almost
all the contemporary developments including Heisenberg's matrix mechanics,
Schr\"{o}dinger's wave mechanics, Born's probability interpretation, and
finally Jordan's and Dirac's transformation theory to unify them into
a single formalism. These lectures were published \cite{key-15},
which was the last publication by Hilbert on the subject possibly
because of his failing health he could not keep up with the fast moving
subject. However, he was thrilled by the marvelous coincidence that
the mathematical formalism developed by him before the advent of quantum
mechanics ideally harmonized with the formulation of the mathematical
apparatus necessary for an elegant and rigorous treatment of the emerging
new subject.

Von Neumann, as the most outstanding of Hilbert's heirs, continued
to carry the torch and in 1927 published three brilliant papers \cite{key-16,key-17,key-18}
that placed quantum mechanics on a meticulous mathematical foundation
including a rigorous proof of the equivalence of matrix and wave mechanics.
He also coined the name Hilbert space that soon took off with every
one starting to use it for the new formalism. The significance of
the concept of a Hilbert space was underlined with the realization
that it offers one of the best mathematical formulations of quantum
mechanics.

His habilitation (qualification to conduct independent university
teaching) was completed in December of 1927, and he began lecturing
as a Privatdozent at the University of Berlin in 1928 at the age of
25, the youngest Privatdozent ever elected in the university's history
in any subject. In 1929, he briefly became a Privatdozent at the University
of Hamburg, where the prospects of becoming a tenured professor were
better, but in October of that year a better offer presented itself
when he was invited to Princeton University. In 1933, he was offered
a lifetime professorship at the Institute for Advanced Study in New
Jersey. He remained a mathematics professor there until his death,
although he had announced his intention to resign and become a professor
at large at the University of California, Los Angeles.

As a true polymath, he made significant contributions to various diverse
fields throughout his momentous career. It is fascinating to note
that he himself considered his pioneering contributions to the foundations
of quantum mechanics especially the Hilbert state formalism as his
most important scientific contribution.

In a recent article, Klaas Landsman \cite{key-19} summarizes von
Neumann's main accomplishments in his pioneering contribution to the
development of the Hilbert state formalism of quantum mechanics, with
the admiring comment that any one of these would have been a significant
achievement for a 23 year old. These are:
\begin{quote}
1. Axiomatization of the notion of a Hilbert space (previously known
only in examples).

2. Establishment of a spectral theorem for (possibly unbounded) self-adjoint
operators.

3. Axiomatization of quantum mechanics in terms of Hilbert spaces
(and operators):

(a) Identification of observables with (possibly unbounded) self-adjoint
operators.\\
(b) Identification of pure states with one-dimensional projections
(or rays).\\
(c) Identification of transition amplitudes with inner products.\\
(d) A formula for the Born rule stating the probability of measurement
outcomes.\\
(e) Identification of general states with density operators.\\
(f) Identification of propositions with closed subspaces (or the projections
thereon). \cite{key-19}
\end{quote}

\subsection{Dirac's contribution}

In the meantime, another star shined in the rising quantum firmament.
Paul Dirac received his PhD in Physics under Ralph Fowler at Cambridge
in 1926 just about the same time did von Neumann. It is quite interesting
to note that both of them acquired a PhD in engineering before veering
off to theoretical physics. Dirac, however, was the first physicist
to ever get a PhD in quantum mechanics, more specifically in matrix
mechanics, which he improved by formally characterizing it using non-commutative
operators and Poisson's brackets.

Schr\"{o}dinger initially revealed the similarity between the two seemingly
different formulations of quantum mechanics, his own wave mechanics
and Heisenberg's matrix mechanics. In early 1927, Dirac \cite{key-20}
and Jordan \cite{key-21,key-21b}, independently of one another, published
their versions of a general formalism tying the various forms of the
new quantum theory together in full generality. This formalism has
come to be known as the Dirac--Jordan transformation theory.

A few months later, in response to these publications by Dirac and
Jordan, John von Neumann published his Hilbert space formalism for
quantum mechanics. Jordan's work in time went into oblivion while
Dirac firmly embraced the Hilbert space formalism advanced by von
Neumann. Additionally Dirac introduced the elegant bra-ket notation
as well as delta functions. Von Neumann did not considered use of
the delta functions to be rigorous. However, a later version oddly
dubbed the rigged Hilbert space was constructed, which restored rigorousness
to Dirac's approach.

The contributions of von Neumann and Dirac to the foundations of quantum
theory using Hilbert space are considered to be equal as reflected
in the embraced phrase the Dirac--von Neumann axioms in mathematical
formulation of quantum mechanics. Compilations of their innovation
were published as books by Paul Dirac \cite{key-22} in 1930 and John
von Neumann \cite{key-23} in 1932. In many ways their contributions
are mutually complementary. For example, while von Neumann's contributions
often emphasized mathematical rigor, Dirac emphasized pragmatic concerns
such as utility and intuitiveness.

\section{Fundamentals of Hilbert Space Formalism}

The Hilbert space, acknowledged as the most appropriate for mathematical
formulations of quantum mechanics, is a square integrable, complex,
linear, abstract space of vectors possessing a positive definite inner
product assured to be a number. The states of a quantum mechanical
system are vectors in a multidimensional Hilbert space containing
an orthonormal basis set of eigenfunctions. The observables are Hermitian
operators on that space, and measurements are orthogonal projections.
Unitary operators are used for changing a vector from one basis set
to another. For elegance, Dirac's bra-ket notation is used to characterize
the vectors and delta functions are used to express orthogonality
of vectors.

No physical property of a quantum system changes by going from wave
or matrix mechanics rendering to the Hilbert state formalism. A very
simple example illustrates the essence. The normalization relation
for a single particle wave packet function in position space is presented
in the wave mechanical description as
\begin{equation}
\int_{-\infty}^{+\infty}\psi^{*}\left(x\right)\psi\left(x\right)dx=1\label{eq:6}
\end{equation}
Using Dirac's bra-ket notation equation (\ref{eq:6}) becomes
\begin{equation}
\int_{-\infty}^{+\infty}\psi^{*}\left(x\right)\psi\left(x\right)dx=\langle\psi|\psi\rangle=1
\end{equation}
where the bra vector $\langle\psi|$ is complex conjugate transpose of the ket~$|\psi\rangle$.

The orthogonality relation is given by
\begin{equation}
\int_{-\infty}^{+\infty}\psi_{m}^{*}\left(x\right)\psi_{n}\left(x\right)dx=\langle\psi_{m}|\psi_{n}\rangle=\delta_{mn}
\end{equation}
where
$\delta_{mn}$ is the Kronecker delta
\begin{equation}
\delta_{mn}=\begin{cases}
0 & \textrm{if }m\neq n\\
1 & \textrm{if }m=n
\end{cases}
\end{equation}
The quantum wave functions, for example, the solutions of the Schr\"{o}dinger
equation describing physical states in wave mechanics are considered
as the set of components $\psi(x)$ of the abstract vector $\Psi$
called the state vector. However, the state vector does not depend
upon any particular choice of coordinates. The same state vector can
be described in terms of the wave function in position or momentum
state or write as an expansion in wave functions $\psi_{n}(x)$ of
definite energy
\begin{equation}
\Psi=\sum_{n}c_{n}\psi_{n}(x)
\end{equation}
suggesting that every linear combination of vectors in a Hilbert space
is again a vector in the Hilbert space. The normalized square moduli
$|c_{n}|^{2}$ of the complex coefficients are then interpreted as
the probability for the system to be in the state $\psi_{n}$ analogous
to Born's initial proposal where $|\psi(x)|^{2}$ is interpreted as
the probability density for the particle to be at $x$.

\subsection{Projective measurement}

Every vector in the Hilbert space as a linear combination of the basis
vectors $\psi_{n}$ with complex coefficients $c_{n}$ can be expressed
in Dirac's notation as
\begin{equation}
|\psi\rangle=\sum_{n}c_{n}|\psi_{n}\rangle.\label{eq:11}
\end{equation}
Multiplying both sides of equation (\ref{eq:11}), by $\langle\psi_{m}|$
gives
\begin{equation}
\langle\psi_{m}|\psi\rangle=\sum_{n}c_{n}\langle\psi_{m}|\psi_{n}\rangle.
\end{equation}
Since $\langle\psi_{m}|\psi_{n}\rangle=\delta_{mn}$, $\langle\psi_{n}|\psi_{n}\rangle=1$
\begin{equation}
c_{n}=\langle\psi_{n}|\psi\rangle\label{eq:13}
\end{equation}
which is the transition amplitude of state $|\psi\rangle$ to state
$|\psi_{n}\rangle$.

Inserting equation (\ref{eq:13}) into equation (\ref{eq:11}) gives
\begin{equation}
|\psi\rangle=\sum_{n}|\psi_{n}\rangle\langle\psi_{n}|\psi\rangle.\label{eq:14}
\end{equation}
Defining a projection operator $\hat{P}_{n}=|\psi_{n}\rangle\langle\psi_{n}|$,
equation (\ref{eq:14}) becomes
\begin{equation}
|\psi\rangle=\sum_{n}\hat{P}_{n}|\psi\rangle.
\end{equation}
leading to
\begin{equation}
\sum_{n}\hat{P}_{n}=\hat{I}
\end{equation}
signifying that the sum
of all the projection operators is unity.

The outer product $|\psi\rangle\langle\psi|$ is called the projection
operator since it projects an input ket vector $|\phi\rangle$ into
a ray defined by the ket $|\psi\rangle$, as follows
\begin{equation}
|\psi\rangle\langle\psi||\phi\rangle=\left(\langle\psi|\phi\rangle\right)|\psi\rangle
\end{equation}
with a probability $|\langle\psi|\phi\rangle|^{2}$ as the inner product
between two state vectors is a complex number known as probability
amplitude. This is usually known as projective measurement and we
will notice that it is important for the measurement of a mixed state
consisting of an ensemble of pure states.

\subsection{Operator Valued Observables}

In a quantum system, what can be measured in an experiment are the
eigenvalues of various observable physical quantities like position,
momentum, energy, etc. These observables are represented by linear,
self-adjoint Hermitian operators acting on Hilbert space.

Each eigenstate of an observable corresponds to eigenvector $|\psi_{n}\rangle$
of the operator $\hat{A}$, and the associated eigenvalue $\lambda_{n}$
corresponds to the value of the observable in that eigenstate
\begin{equation}
\hat{A}|\psi_{n}\rangle=\lambda_{n}|\psi_{n}\rangle\label{eq:17}
\end{equation}
For a self-adjoint Hermitian operator, quantum states associated with
different eigenvalues of $\hat{A}$ are orthogonal to one another
\begin{equation}
\langle\psi_{m}|\psi_{n}\rangle=\delta_{mn}
\end{equation}
The possible results of a measurement are the eigenvalues of the operator,
which explains the choice of self-adjoint operators for all the eigenvalues
to be real. The probability distribution of an observable in a given
state can be found by computing the spectral decomposition of the
corresponding operator. For a Hermitian operator $\hat{A}$ on an
$n$-dimensional Hilbert space, this can be expressed in terms of
its eigenvalues $\lambda_{n}$ following equation (\ref{eq:17}) as
\begin{equation}
\hat{A}=\sum_{n}\lambda_{n}|\psi_{n}\rangle\langle\psi_{n}|
\end{equation}
If the observable $\hat{A}$, with eigenstates $\left\{ |\psi_{n}\rangle\right\} $
and spectrum $\left\{ \lambda_{n}\right\} $ is measured on a system
described by the state vector $|\psi\rangle$, the probability for
the measurement to yield the value $\lambda_{n}$ is given by
\begin{equation}
p(\lambda_{n})=|\langle\psi_{n}|\psi\rangle|^{2}
\end{equation}
After the measurement the system is in the eigenstate $|\psi_{n}\rangle$
corresponding to the eigenvalue $\lambda_{n}$ found in the measurement,
which is called the reduction of state. This seemingly unphysical
reduction of state is a shortcut for the description of the measurement
process and the fact that the system becomes entangled with the state
of the macroscopic measurement equipment. The entanglement leads to
the necessary decoherence of the superposition of states of the measured
system leading solely to the observed eigenvalue with its specific
probability.

\subsection{Unitary Operators}

If the inverse of an operator $\hat{U}$ is the adjoint operator
\begin{equation}
\hat{U}^{-1}=\hat{U}^{\dagger}
\end{equation}
then this operator is called a unitary operator and
\begin{equation}
\hat{U}^{\dagger}\hat{U}=\hat{U}\hat{U}^{\dagger}=\hat{I}
\end{equation}
Unitary operators play a significant role in quantum mechanics representing
transformation in the state space. Time evolution is just one example
since the evolution of state vectors with time is unitary. This means
the state vector changes smoothly preserving the total probability.

\subsection{Quantum Entanglement}

In all of the operations in the Hilbert space formalism, the reality
of the quantum state is not altered. We now consider the landmark
paper \cite{key-24}, of Einstein--Podolsky--Rosen (EPR) where they
presented a thought experiment, which attempted to show that ``the
quantum-mechanical description of physical reality given by wave functions
is not complete'', indicating the possible existence of some hidden
variables to explain the apparent violation of locality enshrined
in Einstein's theory of relativity. However, in 1964, John Stewart
Bell offered \cite{key-2} his celebrated theoretical explanation
known as Bell's inequality revealing that one of the key assumptions,
the principle of locality, as applied to the kind of hidden variables
interpretation hoped for by EPR, was mathematically inconsistent with
the predictions of quantum theory.

Among an overabundance of experimental efforts performed under the
generic topic of quantum entanglement, a loophole-free Bell inequality
violation has presumably claimed to have been conclusively demonstrated
\cite{key-25}. These experiments demonstrate that although there
are some possible non-local correlation in quantum systems, it does
not violate causality since no information can be transferred faster
than the speed of light consistent with the theory of relativity.

A detailed discussion has been provided by Bhaumik \cite{key-26}
illustrating how the reality of the quantum state is not violated
by quantum entanglement. In brief, the expectation value of an overall
pure quantum state of a composite system does not change although
in an entanglement experiment, wave function of a constituent mixed
subsystem can change violating locality.

This is due to the fact that the actions of an experimentalist on
a subsystem of an entangled state can be described as applying a unitary
operator to that subsystem. Although this produces a change on the
wave function of the complete system, such a unitary operator cannot
change the density matrix describing the rest of the system. In brief,
if distant particles 1 and 2 are in an entangled state, nothing an
experimentalist with access only to particle 1 can do that would change
the density matrix of particle 2.

The density matrix $\hat{\rho}$ of an ensemble of states $|n\rangle$
with probabilities $p_{n}$ is given by
\begin{equation}
\hat{\rho}=\sum_{n}p_{n}|n\rangle\langle n|
\end{equation}
where $|n\rangle\langle n|$ are projection operators and the sum
of the probabilities are $\sum_{n}p_{n}=1$. Thus there can be various
ensembles of states with each one having its own probability distribution
that will give the same density matrix.

The expectation value of an observable $\langle\hat{A}\rangle$ is
\begin{equation}
\langle\hat{A}\rangle=\textrm{Tr}\left(\hat{\rho}\hat{A}\right)
\end{equation}
Furthermore, the time evolution of $\hat{\rho}$ only depends upon the commutator of $\hat{\rho}$ with the Hamiltonian $\hat{H}$
following the von Neumann equation
\begin{equation}
\imath\hbar\frac{d}{dt}\hat{\rho}(t)=[\hat{H},\hat{\rho}(t)]
\end{equation}

Thus as long as $\hat{\rho}$ remains the same, a change in the wave
function of particle 2 does not affect any observable since all observable
results can be predicted from the density matrix, without needing
to know the ensemble used to construct it. Consequently no useful
signal can be sent using entanglement and nonlocality between two
observers separated by an arbitrary distance thereby no violation
of the sanctified tenets of special theory of relativity ensues.

\section{Conclusion}

It would be cogent to acknowledge the space filling, ever immutable,
universal quantum fields to constitute the primary ingredients of
reality uncovered by us so far. Elementary particles like electrons
are quanta of these fields and as such are as real as the fields themselves.
And so is the wave packet functions depicting elementary particles
constructed in terms of the attributes of the quantum fields.

The quantum states of, for example, a hydrogen atom portrayed in terms
of the wave functions that are solutions of the Schr\"{o}dinger equation
using the wave packet function ought to be real as well. In fact,
recent experimental observations confirm this reality. It is now well
known that the same quantum states can also be described in terms
of Heisenberg's matrix mechanics. Eventually it was recognized that
the most elegant and efficient way to treat the quantum states is
by utilizing the abstract Hilbert space formalism developed predominantly
by John von Neumann and Paul Dirac.

There have been concerns about whether the quantum states described
by the abstract Hilbert space have any reality. Using appropriate
assumptions several recent theoretical treatments suggest that the
quantum states in Hilbert space are real. However, the somewhat roundabout
ways of going about the proof leave rooms for possible loopholes.
We show here that the reality of the quantum states can be confirmed
in a straightforward manner relying on the primary reality of quantum
fields.

\section*{Acknowledgements}

The author wishes to thank Zvi Bern, Per Kraus, and Eric D'Hoker for
helpful discussions.

\pagebreak
\balance

\end{document}